\documentclass[12pt,preprint]{aastex}

\begin{document}

\title{The Maryland-NOAO Instrument Partnership (2003-2009)}

\author{Sylvain Veilleux}
\affil{Department of Astronomy, University of Maryland,
  College Park, MD 20742; veilleux@astro.umd.edu}

\author{Buell T. Jannuzi}
\affil{National Optical Astronomy Observatory, 950 N. Cherry Ave,
  Tucson, AZ 85726; jannuzi@noao.edu}

Seven years ago, with the encouragement of the NSF and AURA, NOAO
requested proposals from the community to partner with the national
observatory to improve instrumentation and/or telescope capabilities
at KPNO and CTIO.  Of the proposals that were selected, one came from
the University of Maryland with the goals of helping NOAO complete the
development, construction, and deployment of a new, wide-field,
near-IR imager (NEWFIRM) and of working with NOAO to develop data
reduction pipelines and data archiving capabilities at NOAO.

By all measures, the Maryland-NOAO instrument partnership has been a
resounding success.  NOAO and the astronomical community have
benefited from the use of Maryland resources and staff to support
the development of NEWFIRM; the One Degree Imager (ODI), an
innovative, wide-field optical imager; and the NOAO Science Archive
(NSA).  University of Maryland funds also were used to help support
NOAO’s share of the costs of constructing ODI for WIYN.  The
University of Maryland procured key hardware components for NEWFIRM
and acquired a suite of narrowband filters that were not affordable in
the baseline budget.  Maryland personnel Rob Swaters, Tracy Huard, and
Brian Thomas became fully integrated into the NOAO pipeline and NSA
development teams.  Their participation made the critical difference
for meeting the delivery schedule of the quick-reduce pipeline with
the commissioning of NEWFIRM.  They helped with, and often led, the
design, implementation, and testing of several aspects of the NOAO
High-Performance Pipeline System (NHPPS) and the NSA.

The NHPPS, now in routine operation, is used to process data obtained
with Mosaic and NEWFIRM on the 4-meter telescopes at KPNO and CTIO.
The data processing carried out by NHPPS consists of two components: a
Quick-Reduce Pipeline that reduces data in near real time at the
telescope and a Science Pipeline that provides in-depth data reduction
at the end of an observing block.  Examples of NEWFIRM data processed
by NHPPS were discussed in the December 2008 NOAO/NSO Newsletter
(\#96, pp. 1 and 21–22; Gutermuth \& Dickinson).

From the point of view of Maryland, this partnership has had a direct,
positive impact on the quantity and quality of the science produced by
members of the Department of Astronomy.  It allowed Maryland
astronomers and students to (1) explore new avenues of research
including high-reward projects that would perhaps have been considered
too risky by a national Time Allocation Committee, (2) quickly respond
to new discoveries and targets of opportunity, and (3) undertake large
thesis and non-thesis surveys with a guarantee of telescope time.

Under this partnership, 11 professorial faculty, 11 research faculty
(i.e., post-docs and research scientists), 16 graduate students
(including 13 PhD theses), and 3 undergraduate students from Maryland
used the KPNO facilities.  Thirteen papers directly involving KPNO
data have been published so far under this partnership (11 of these
papers have Maryland graduate students as first authors); many more
are in preparation.  A necessarily incomplete list of scientific
highlights includes the successful monitoring of comet Tempel 1
between February and July 2005 in support of the Deep Impact mission
(M.  A’Hearn, PI), an extension of the ``Cores to Disks'' Spitzer Legacy
Program to shorter, near-infrared wavelengths (S. Chapman, PhD thesis;
L. Mundy, advisor), a study of the density profiles of the dark matter
halos in low-surface brightness galaxies and low-mass dwarf galaxies
(R. Kuzio de Naray, PhD thesis; S. McGaugh, advisor), and a
spectroscopic survey for superwinds in massive starbursts (D. Rupke,
PhD thesis; S. Veilleux, advisor).

he Maryland-NOAO partnership also has had a direct positive impact on
the technical capability of Maryland’s Department of Astronomy.  It
helped strengthen the Department’s software expertise in the areas of
optical and near-infrared data reduction and archiving, complementing
the already strong software group at millimetric wavelengths
associated with the Combined Array for Research in Millimeter-wave
Astronomy (CARMA) collaboration.  The lead role assumed by the
Maryland personnel in the development of NHPPS and NSA helped create
an in-house resource group with the highest level of expertise capable
of advising Maryland users and collaborators, as well as members of
the astronomical community-at-large, on all aspects of the analysis of
Mosaic and NEWFIRM data.  It also helped position the Maryland
software group for similar projects at national or private
observatories in the future.

The Maryland-NOAO partnership increased the visibility of the
Department within the College and the University.  It was branded by
the administration as an excellent example of successful collaboration
between a University and a national laboratory.  No doubt it will be
regarded as a model to follow in future partnerships in which the
University of Maryland becomes engaged.  For KPNO, the partnership
helped keep the Mayall a competitive, modern facility by helping to
ensure the arrival of NEWFIRM.  Maryland’s critical involvement in the
development of the first ``real-time'' pipelines in use for our
instruments has helped continue the development of the modern science
capabilities of KPNO.  Finally, the returning users of our facilities
from Maryland provided valuable, close ties to a segment of our user
community and valuable feedback and suggestions from experienced users
of our facility.  If and when we enter into new partnerships, we will
endeavor to repeat the success of the Maryland-NOAO partnership.

\clearpage

\begin{figure*}
\epsscale{1.05}
\plotone{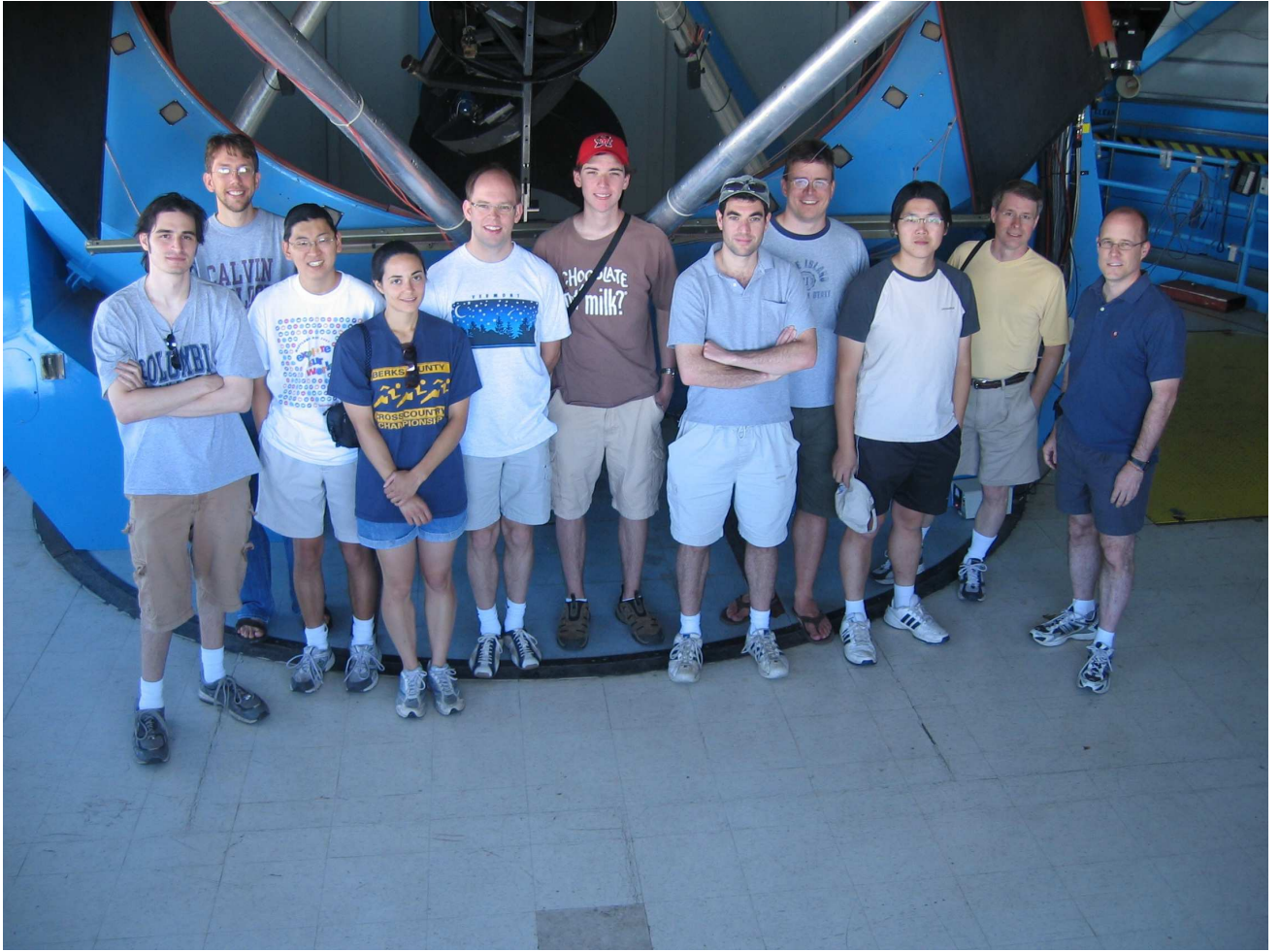}
\caption{Participants and instructors of the 2007 Kitt Peak Summer School.}
\end{figure*}

\begin{figure*}
\epsscale{0.8}
\plotone{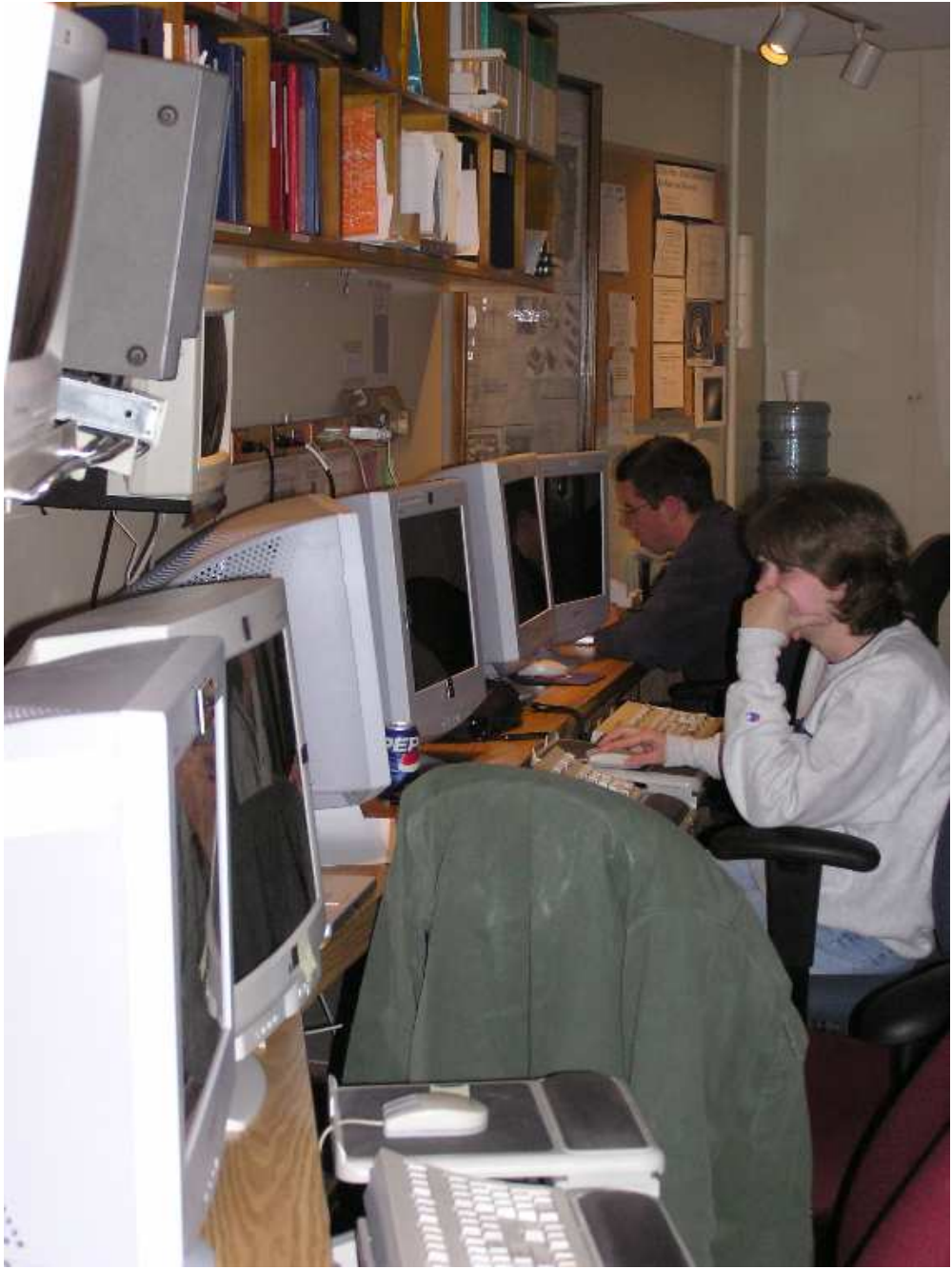}
\caption{Maryland graduate student Rachel Kuzio de Naray (foreground)
  in the Mayall 4-meter control room.}
\end{figure*}

\end{document}